\documentclass{elsart}
\usepackage{graphicx,amssymb}
\journal{Physics Letters B}
\begin{document}
\addtolength{\topmargin}{+50pt}

\begin{frontmatter}

\title{Spacetime structure of massive gravitino}

\author{M. Kirchbach},
\ead{kirchbach@chiral.reduaz.mx}
\author{D. V. Ahluwalia \corauthref{cor}}
\ead{ahluwalia@phases.reduaz.mx}
\corauth[cor]{Corresponding author.}
\address{Fac. de Fisica de la UAZ, Zacatecas, Ap. Postal C-600,
ZAC 98062, Mexico}

\begin{abstract}

We present reasons as to why an {\em ab initio} analysis of 
the spacetime structure of massive gravitino is necessary. Afterwards, 
we construct the relevant representation space, and
finally, give a new physical interpretation of massive gravitino.


\end{abstract}
\end{frontmatter}

\centerline{{\bf Journal-ref:} Phys. Lett. B 529 (2002) 124-131.}

\def\beq{\begin{eqnarray}}
\def\eeq{\end{eqnarray}}

\def\p{(\vec p\,)}
\def\A{{\mathcal A}^\mu}
\def\W{{\mathcal W}_\mu}


\section{Motivation} 
A massive gravitino is described by 
$\psi^\mu$. 
As far as its spacetime properties are
concerned, it transforms as a finite dimensional non-unitary 
representation of the Lorentz group,
\beq
\psi^\mu:\quad
\underbrace{
\left[\left(\frac{1}{2},0\right)\oplus\left(0,\frac{1}{2}\right)\right]
}_{\mbox{\sc Spinor}\;\mbox{\sc Sector}}
	\otimes
\underbrace{
\left(\frac{1}{2},\frac{1}{2}\right)
}_{\mbox{\sc Vector}\;\mbox{\sc Sector}}\,. \label{rs}
\eeq
The unitarily transforming physical states are built upon 
this structure \cite{sw}.

We enumerate two circumstances that  motivate us to take 
an {\em ab initio\/} look at this representation space.

\begin{enumerate}

\item
For the vector sector, it has recently  been called to
attention  that
the Proca description of the $(1/2,1/2)$ representation space is
incomplete \cite{ak}. An {\em ab initio\/} construction of this 
sector reveals that the St\"uckelberg contribution to the propagator, 
so important for the renormalization of the gauge theories with 
massive bosons \cite{v}, is found to naturally reside in the   
$(1/2,1/2)$ representation space.

\item
At the same time, the properties of the
$(1/2,1/2)$, along with that of 
the $(1/2,0)\oplus(0,1/2)$, representation space determine the
structure of
$\psi^\mu $. In order to impose a single-spin, i.e., spin $3/2$, 
interpretation on the latter, the lower spin-$1/2^+$ and
spin-$1/2^-$ components of $\psi^\mu$ are considered as redundant, 
unphysical, states that are claimed to be excluded from consideration 
by means of the two supplementary conditions:
$\gamma_\mu\psi^\mu(x)=0$, and $\partial_\mu \psi^\mu(x)=0$,
respectively. However, this time-honored 
framework was questioned by a recent empirical observation
regarding the $N$  and $\Delta $ resonances \cite{mk}.
The available data on high-spin resonances
reveal an unexpected and systematic clustering 
in terms of the $(j/2, j/2)\otimes \lbrack (1/2,0)\oplus (0,1/2)\rbrack $
representations with $j=1,3$ and $5$  {\em without\/} imposition of the 
supplementary conditions. 
For the $N$ and $\Delta$ resonances these results are summarized in
Fig. 1. For example,  in the standard theoretical 
framework $N(1440)$, $N(1535)$, $\Delta(1620)$ and  $\Delta(1750)$ 
should have been absent.  Experimental data shows them to be 
present at statistically significant level.\footnote{The $N(1440)$, $N(1535)$, 
$\Delta(1620)$ carry four star status, while at present  $\Delta(1750)$ 
simply has a one star significance \cite{pdg}.}

\end{enumerate}

In regard to the latter of the two enumerated circumstances, we take
the position that any solution of the  QCD Lagrangian for particle 
resonances must carry well-defined transformation properties when looked
upon from different inertial frames. This forces these resonances to 
belong to one, or the other, of various representation spaces of the
Lorentz group. For this reason the data on particle resonances 
may furnish hints on physical interpretation of various 
Lorentz group representations that one needs in gauge theories, or theories
of supergravity.

For exploring the spacetime structure of massive gravitinos
the charge conjugation properties play an important role.
Under the operation of charge conjugation, 
one may choose the spinor sector to behave as a Dirac object, and implement
the Majorana nature of the massive gravitino at the level of the
Fock space. This is standard, see, e.g., Ref. \cite{Moroi}. Or, from the
very beginning choose the spinor sector to behave as a Majorana object.
Since we wish to stress certain non-trivial aspects
of massive gravitino that do not -- at least qualitatively -- depend 
on this choice, we shall here treat the spinor sector to be of Dirac 
type. 

Very nature of our   {\em ab initio} look at the representation
space defined in Eq. (\ref{rs}), obliges us to present 
sufficient pedagogic details so that by the end of the
paper much that is needed to form an opinion on the arrived results
is readily available. At the same time, {\em Letter\/} nature of this
manuscript would prevent us from delving into subtle details which are,
for present, of secondary importance (but have been studied and are
planned to be presented elsewhere).

We shall work in the momentum space.
The notation will be essentially that introduced in Ref. \cite{ak}.

\section{Construction of the Spinor and Vector Sectors}
\label{sect: sv}

We now wish to construct the primitive objects that span the 
representation space 
defined in Eq. (\ref{rs}) for an arbitrary 
$\vec{p}$. We first construct the spinor and vector 
sectors in the rest frame, and then  boost them using the following
boosts:
\beq
\kappa^{\left(\frac{1}{2},0\right)\oplus\left(0,\frac{1}{2}\right)
}
=
\kappa^{\left(\frac{1}{2},0\right)}
\oplus
\kappa^{\left(0,\frac{1}{2}\right)}\,,\quad
\kappa^{\left(\frac{1}{2},\frac{1}{2}\right)}
=
\kappa^{\left(\frac{1}{2},0\right)} 
\otimes
\kappa^{\left(0,\frac{1}{2}\right)}\,,
\eeq
with
\beq
\kappa^{\left(\frac{1}{2},0\right)}&=&
\frac{1}{\sqrt{2m(E+m)}}\left[(E+m)I_2+\vec\sigma\cdot\vec p\,\right]\,,\\
\kappa^{\left(0,\frac{1}{2}\right)}&=&
\frac{1}{\sqrt{2m(E+m)}}\left[(E+m)I_2-\vec\sigma\cdot\vec p\,\right]\,.
\eeq
where $I_n$ stands for an $n\times n$ identity matrix,
while the remaining  symbols carry their usual contextual meaning. 
We define the spin-$1/2$
helicity operator:
$
\Sigma=({\vec \sigma}/{2})\cdot\widehat{p},
$
where $\widehat{p} = \vec{p}/\vert \vec{p}\,\vert$, and
$\vec p=\vert \vec{p}\,\vert ( \sin(\theta)\cos(\phi), 
 \sin(\theta)\sin(\phi),
\cos(\theta))$.
Keeping full freedom in the choice of phases, its positive and negative
helicity states are:
\beq
&& h^+= N \exp\left(i \vartheta_1\right)
\left(
\begin{array}{c}
\cos(\theta/2) \exp(-i \phi/2)\\
\sin(\theta/2) \exp(i \phi/2)
\end{array}
\right),\nonumber\\
&& h^-= N \exp\left(i \vartheta_2\right)
\left(
\begin{array}{c}
\sin(\theta/2) \exp(-i \phi/2)\\
-\cos(\theta/2) \exp(i \phi/2)
\end{array}
\right)\,.\label{h}
\eeq

\subsection{$(1/2,0)\oplus(0,1/2)$ Representation space}
\label{2.1}

In this subsection we present the kinematic structure 
of the $(1/2,0)\oplus(0,1/2)$ representation space in such a manner
that its extension to the $(1/2,1/2)$  representation space becomes
transparent. The familiarity of the equation presented shall
not, we hope, mar the procedure adopted. The entire construct, in
essence, relies on nothing more than the boost operators.

The rest-frame $(1/2,0)\oplus(0,1/2)$ spinors are then chosen to be:
\beq
&& u_{+1/2}(\vec 0\,)=
\left(
\begin{array}{c}
h^+\\
h^+
\end{array}
\right)\,,\quad u_{-1/2}(\vec 0\,)=
\left(
\begin{array}{c}
h^-\\
h^-
\end{array}
\right)\,,\nonumber\\
&& v_{+1/2}(\vec 0\,)=
\left(
\begin{array}{c}
h^+\\
-h^+
\end{array}
\right)\,,\quad
v_{-1/2}(\vec 0\,)=
\left(
\begin{array}{c}
h^-\\
-h^-
\end{array}
\right).
\eeq
The choice of the phases made in writing down these spinors
has been determined by the demand of parity covariance.

The 
boosted spinors, $u_{\pm 1/2}(\vec p\,)$ and $v_{\pm 1/2}(\vec p\,)$
are obtained by applying the boost operator
$\kappa^{\left(\frac{1}{2},0\right)\oplus\left(0,\frac{1}{2}\right)
}$ to the above spinors, yielding: 
\beq
&& u_{+1/2}(\vec p\,)=
\frac{N \exp(i \vartheta_1)}{\sqrt{2 m ( m +E)}}
\left(
\begin{array}{c}
\exp(-i \phi/2) (m+\vert \vec p\,\vert +E) \cos(\theta/2)\\
\exp(i \phi/2) (m+\vert \vec p\,\vert +E) \sin(\theta/2)\\
\exp(-i \phi/2) (m-\vert \vec p\,\vert +E) \cos(\theta/2)\\
\exp(i \phi/2) (m-\vert \vec p\,\vert +E) \sin(\theta/2)\\
\end{array}
\right)\,,\nonumber\\
&& u_{-1/2}(\vec p\,)=
\frac{N \exp(i \vartheta_2)}{\sqrt{2 m ( m +E)}}
\left(
\begin{array}{c}
\exp(-i \phi/2) (m-\vert \vec p\,\vert +E) \sin(\theta/2)\\
-\exp(i \phi/2) (m-\vert \vec p\,\vert +E) \cos(\theta/2)\\
\exp(-i \phi/2) (m+\vert \vec p\,\vert +E) \sin(\theta/2)\\
-\exp(i \phi/2) (m+\vert \vec p\,\vert +E) \cos(\theta/2)\\
\end{array}
\right)\,,\nonumber\\
&& v_{+1/2}(\vec p\,)=
\frac{N \exp(i \vartheta_1)}{\sqrt{2 m ( m +E)}}
\left(
\begin{array}{c}
\exp(-i \phi/2) (m+\vert \vec p\,\vert +E) \cos(\theta/2)\\
\exp(i \phi/2) (m+\vert \vec p\,\vert +E) \sin(\theta/2)\\
-\exp(-i \phi/2) (m-\vert \vec p\,\vert +E) \cos(\theta/2)\\
-\exp(i \phi/2) (m-\vert \vec p\,\vert +E) \sin(\theta/2)\\
\end{array}
\right)\,,\nonumber\\
&& v_{-1/2}(\vec p\,)=
\frac{N \exp(i \vartheta_2)}{\sqrt{2 m ( m +E)}}
\left(
\begin{array}{c}
\exp(-i \phi/2) (m-\vert \vec p\,\vert +E) \sin(\theta/2)\\
-\exp(i \phi/2) (m-\vert \vec p\,\vert +E) \cos(\theta/2)\\
-\exp(-i \phi/2) (m+\vert \vec p\,\vert +E) \sin(\theta/2)\\
\exp(i \phi/2) (m+\vert \vec p\,\vert +E) \cos(\theta/2)\\
\end{array}
\right)\,.\label{uv}
\eeq
In the standard notation, these satisfy the orthonormality and
completeness  relations, along with the Dirac equation:
\beq
&& \bar u_h(\vec p\,)\, u_{h^\prime}(\vec p\,) =  +2 N^2
 \delta_{h {h^\prime}}\,,
\quad
\bar v_h(\vec p\,)\, v_{h^\prime}(\vec p\,) =  -2 N^2
 \delta_{h {h^\prime}}\,,\label{5}\\
&& \frac{1}{2 N^2}
\left[\sum_{h=\pm 1/2}
 u_{h}(\vec p\,) \bar u_h(\vec p\,) -
 \sum_{h=\pm 1/2} v_{h}(\vec p\,) \bar v_h(\vec p\,)\right] = I_4\,.\label{1}
\eeq
The wave equation satisfied by the $u_h(\vec p\,)$ and 
$v_h(\vec p\,)$ spinors follows if we note:\footnote{This part of the calculation
is best done if one starts with the momenta in the Cartesian co-ordinates, 
and takes the ``axis of spin quantization'' to be the 
z-axis.}
\beq
\frac{1}{2 N^2}
\left[\sum_{h=\pm 1/2}
 u_{h}(\vec p\,) \bar u_h(\vec p\,) +
 \sum_{h=\pm 1/2} v_{h}(\vec p\,) \bar v_h(\vec p\,)\right]\equiv 
\frac{\gamma_\mu p^\mu }{m}\,,\label{2}
\eeq
where we defined,
\beq
\gamma_0=\left(\begin{array}{cc}
                0_2 & I_2 \\
                I_2 & 0_2
	\end{array}\right)\,,\quad
\gamma_i=\left(\begin{array}{cc}
                0_2 & \sigma_i \\
                -\sigma_i & 0_2
	\end{array}\right)\,.
\eeq
The $0_n$ are $n\times n$ null matrices.
Adding/subtracting Eqs. (\ref{1}) and (\ref{2}), yields:
\beq
\frac{1}{2 N^2}
\sum_{h=\pm 1/2}
 u_{h}(\vec p\,) \bar u_h(\vec p\,) = 
\frac{\gamma_\mu p^\mu + m I_4}{2 m}\,,\label{3}\\
\frac{1}{2 N^2}
\sum_{h=\pm 1/2} v_{h}(\vec p\,) \bar v_h(\vec p\,)
=
\frac{\gamma_\mu p^\mu - m I_4}{2 m}\,.\label{4}
\eeq
Multiplying Eq. (\ref{3}) from the right by $u_{h^\prime}(\vec p\,)$, and 
Eq. (\ref{4}) by $v_{h^\prime}(\vec p\,)$, and using Eqs. (\ref{5}),
immediately yield the momentum-space wave equation for the
$(1/2,0)\oplus(0,1/2)$ representation space,
\beq
\left(\gamma_\mu p^\mu \pm m I_4\right)\psi_h(\vec p\,)=0\,.\label{deq}
\eeq
In Eq. (\ref{deq}), the minus sign
is to be taken for, $\psi_h(\vec p\,) =  u_{h}(\vec p\,)$, and
the plus sign for, $\psi_h(\vec p\,) =  v_{h}(\vec p\,)$.

The essential element to note in regard to Eq. (\ref{deq}) is that 
it follows directly from the explicit expressions for the
$u_h(\vec p\,)$ and $v_h(\vec p\,)$.

\subsection{$(1/2,1/2)$ Representation space -- An ab initio construct}

Next, we introduce the rest-frame vectors for the $(1/2,1/2)$ 
representation space,
\beq
&& \xi_1(\vec 0\,) = h^+\otimes h^+\,,\\
&& \xi_2(\vec 0\,) = 
\frac{1}{\sqrt{2}}\left( h^+\otimes h^- + h^- \otimes h^+\right) \,,\\
&& \xi_3(\vec 0\,) = h^-\otimes h^-\,,\\
&& \xi_4(\vec 0\,) = 
\frac{1}{\sqrt{2}}\left( h^+\otimes h^- - h^- \otimes h^+\right) \,.
\eeq
The boosted vectors are thus: 
$
\xi_\zeta(\vec p\,)= \kappa^{\left(\frac{1}{2},\frac{1}{2}\right)}
\xi_\zeta(\vec 0\,), \zeta=1,2,3,4,
$
\beq
&& \xi_1(\vec p\,)
= \frac{\exp(i 2 \vartheta_1) N^2}{2} 
\left(
\begin{array}{c}
2 \exp(-i \phi) \cos^2(\theta/2) \\
\sin(\theta) \\
\sin(\theta)  \\
2 \exp(i \phi) \sin^2(\theta/2)
\end{array}
\right)\,,\nonumber\\
&& \xi_2(\vec p\,)
= \frac{\exp(i 2 \vartheta_1) N^2}{\sqrt{2} m } 
\left(
\begin{array}{c}
 \exp(-i \phi) E \sin(\theta) \\
- \left(\vert \vec p \,\vert + E \cos(\theta)\right) \\
\vert \vec p \,\vert - E \cos(\theta) \\
- \exp(i \phi) E \sin(\theta)
\end{array}
\right)\,,\nonumber\\
&& \xi_3(\vec p\,)
= \frac{\exp(i 2 \vartheta_1) N^2}{2} 
\left(
\begin{array}{c}
2 \exp(-i \phi) \sin^2(\theta/2) \\
- \sin(\theta) \\
- \sin(\theta)  \\
2 \exp(i \phi) \cos^2(\theta/2)
\end{array}
\right)\,,\nonumber\\
&& \xi_4(\vec p\,)
= \frac{\exp(i 2 \vartheta_1) N^2}{\sqrt{2} m } 
\left(
\begin{array}{c}
 \exp(-i \phi) \vert \vec p\,\vert  \sin(\theta) \\
- \left( E + \vert \vec p \,\vert \cos(\theta)\right) \\
E - \vert \vec p \,\vert  \cos(\theta) \\
- \exp(i \phi) \vert \vec p\,\vert  \sin(\theta)
\end{array}
\right)\,.\label{xi}
\eeq

In the notation of Ref. 
\cite{ak},\footnote{In the cited work 
we presented the $(1/2,1/2)$ representation
space in its parity realization.  Here, the  presentation
is in terms of helicity realization. The two descriptions have mathematically 
similar but physically
distinct structures, which, e.g., show up
in their different behavior under the operation of 
Parity.} these satisfy the orthonormality and
completeness  relations, along with a new wave equation.
The orthonormality and
completeness  relations are:
\beq
&& \overline{\xi}_\zeta(\vec p\,) 
\xi_{\zeta^\prime}(\vec p\,) = -\,N^4\,  \delta_{\zeta\zeta^\prime}
\,,\quad  \zeta=1,2,3
\,,\nonumber \\
&& \overline{\xi}_\zeta(\vec p\,) \xi_{\zeta^\prime}(\vec p\,) = 
+\,N^4\,\delta_{\zeta\zeta^\prime}\,,\quad \zeta=4\,,\nonumber\\
&& \frac{1}{N^4}
\left[   \xi_4(\vec p\,) \overline{\xi}_4(\vec p\,)
- \sum_{\zeta=1,2,3}
\xi_\zeta(\vec p\,) \overline{\xi}_\zeta(\vec p\,)\right] = I_4\,,
\eeq
where
\beq
\overline{\xi}_\zeta(\vec p\,) \equiv \xi_\zeta(\vec p\,)^\dagger 
\lambda_{00}\,,
\eeq
with 
\beq
\lambda_{00} = \left(
\begin{array}{cccc}
-1 & 0 & 0 & 0 \\
0 & 0 & -1 & 0 \\
0 & -1 & 0 & 0\\
0 & 0 & 0 & -1
\end{array}
\right)\,.
\eeq

The parity operator for the $(1/2,1/2)$ representation space is:
\beq
{\mathcal P} = \lambda_{00}\, \exp[i \alpha]\,
\,{\mathcal R}\,,\quad && {\mathcal R}:
 \{\theta \rightarrow
\pi-\theta,  \phi\rightarrow \pi + \phi\}\nonumber\\
&& \alpha = \mbox{a real number}\,,
\eeq
while the helicity operator for this space is, $\vec J\cdot\widehat{p}$,
with $\vec J$ given by:
\beq
J_x=\frac{1}{2}
\left(
\begin{array}{cccc}
0 & 1 & 1  & 0 \\
1 & 0 & 0  & 1\\
1 & 0 & 0  & 1\\
0 & 1  &1  & 0 
\end{array}
\right),\,
J_y=\frac{1}{2}
\left(
\begin{array}{cccc}
0 & -i  & -i  & 0 \\
i &  0  &  0  & -i \\
i &  0  &  0  & -i\\
0 &  i  &  i  &  0 
\end{array}
\right),\,
J_z=
\left(
\begin{array}{cccc}
1 & 0 & 0 & 0 \\
0  & 0  & 0  & 0 \\
0 &  0 & 0 &  0\\
0 & 0 & 0 & -1 
\end{array}
\right)\,.
\eeq

\vskip 1cm
We now must take a small definitional detour towards
the notion of the dragged Casimirs for spacetime symmetries. 
It arises in the following fashion. The  second Casimir operator, $C_2$, 
of the Poincar\'e group is defined as the square of the Pauli-Lubanski
pseudovector:
\beq
{\mathcal W}^\mu = \frac{1}{2}\epsilon^{\mu\nu\rho\sigma} M_{\nu\rho} 
P_\sigma\,,
\eeq	
where $\epsilon^{\mu\nu\rho\sigma}$ is the standard Levi-Civita symbol 
in four dimensions, while $M_{\mu\nu}$ denote generators of the Lorentz
group,
\beq
M_{0i}= K_i\,,\quad M_{ij}= \epsilon_{ijk} J^k\,,
\eeq
where each of the $i,j,k$ runs over $1,2,3$.
The $P_\mu$ are generators of the spacetime translations. In general, these
have non-vanishing commutators with $M_{\mu\nu}$,
\beq
[P_\mu,\, M_{\rho\sigma}] = i 
\left(\eta_{\mu\rho} P_\sigma - \eta_{\mu\sigma} P_\rho\right)\,.\label{6}
\eeq
On using Eq. (\ref{6}), we rewrite $C_2$ as 
\beq
C_2 &=& \frac{1}{4}
\epsilon^{\mu\nu\rho\sigma} 
\epsilon_{\mu\lambda\kappa\zeta} M_{\nu\rho}
M^{\lambda\kappa} P_\sigma P^\zeta
\nonumber\\ 
&& \qquad+\left[\frac{i}{4}
\epsilon^{\mu\nu\rho\sigma} \epsilon_{\mu\lambda\kappa\zeta}
M_{\nu\rho} {\eta_\sigma}^\lambda P^\kappa P^\zeta
+
\frac{i}{4}
\epsilon^{\mu\nu\rho\sigma} \epsilon_{\mu\kappa\lambda\zeta}
M_{\nu\rho}
{\eta_\sigma}^\kappa P^\lambda P^\zeta\right]\,.
\label{drg}
\eeq

\vskip 1cm
The squared brackets vanishes due to antisymmetry 
of the Levi-Civita symbol. As such, the space-time translation operators
entering the definition of $C_2$ can be moved to the very right. 

This observation  allows for introducing  the dragged Casimir 
$\widetilde{C}_2$ as an operator with the same  
form as $C_2$ --  the difference being that 
the commutator in Eq.~(\ref{6}) 
is now set to zero [as is appropriate for finite dimensional
$SU_R(2)\otimes SU_L(2)$ representations]. Consequently,
while $C_2$ and $\widetilde{C}_2$ carry same invariant eigenvalues when
acting upon momentum eigenstates, their commutators with the Lorentz group
generators are no longer identical.\footnote{To avoid confusion, note that 
$\widetilde{C}_2$ is defined in the $SU_R(2)\otimes SU_L(2)$; while
$C_2$ is defined in the Poincar\'e group.}
For the $(1/2,0)\oplus(0,1/2)$ representation space,
$[\widetilde{C_2},\,\vec J^{\,2}] $
 vanishes. 
For the $(1/2,1/2)$ representation space,
$[\widetilde{C_2},\,\vec J^{\,2}] $
does not vanish (except when acting upon rest states), 
and equals  $ -\,4\,E\, \vec P\cdot \vec{K}$. This leads to
the fact that while the former representation space is endowed with
a well-defined spin, the latter is not:

\begin{quote}

As an immediate application, $\widetilde{C}_2$ for the 
 $(1/2,1/2)$ representation space
bifurcates this space into two sectors.
The three states $\xi_\zeta(\vec p\,) $ with $\zeta=1,2,3$ 
are associated with the $\widetilde{C}_2$ eigenvalue, 
$-\, 2\, m^2$; while
the, $\zeta=4$, corresponds to eigenvalue zero.

Thus, all the $\xi_\zeta(\vec p\,)$, except for the rest frame, cease 
to be eigenstates of the $(1/2,1/2)$'s $\vec J^{\,2}$ and 
do not carry definite spins.
This contrasts with the situation for the $(1/2,0)\oplus(0,1/2)$ 
representation space, where the $\psi_h(\vec p)$ are eigenstates of the 
corresponding $\vec J^{\,2}$. 

\end{quote}

\vskip 1cm


Now in  order that the  $\xi_\zeta(\vec p\,)$  carry the standard
contravariant Lorentz index, we introduce
a rotation in the $(1/2,1/2)$ representation space via
 \cite{ak}:
\beq
S=\frac{1}{\sqrt{2}}
\left(
\begin{array}{cccc} 
0 & i & -i & 0 \\
-i & 0 & 0 & i \\
1 & 0 & 0 & 1 \\
0 & i & i & 0
\end{array}
\right)
\eeq
Then, the $(1/2,1/2)$ representation space is spanned by four Lorentz 
vectors:
\beq
{\mathcal A}^\mu_\zeta(\vec p\,)= 
S^{\mu\alpha} \left[\xi_\zeta(\vec p\,)\right]_\alpha
\,, \quad\zeta=1,2,3,4\,,
\eeq
and the superscript $\mu$ is the standard Lorentz index.
Following the procedure established 
in Sec. \ref{2.1}, they can be shown to 
satisfy a new wave equation \cite{ak},
\beq
\left(\Lambda_{\mu\nu} 
 p^\mu p ^\nu \pm m^2 I_4\right) {\mathcal A}_\zeta(\vec p\,)=0\,, 
\eeq
where the plus sign is to be taken for, $\zeta=1,2,3$, while the
minus sign belongs to, $\zeta=4$.
The $\Lambda_{\mu\nu}$ 
matrices are: $\Lambda_{00}=\mbox{diag}(1,-1,-1,-1)$,
$\Lambda_{11}=\mbox{diag}(1,-1,1,1)$, $\Lambda_{22}=\mbox{diag}(1,1,-1,1)$,
$\Lambda_{33}=\mbox{diag}(1,1,1,-1)$, and 
\beq
&& \Lambda_{01}=
\left(\begin{array}{cccc}
0 & -1 & 0 & 0 \\
1 & 0 & 0 & 0 \\
0 & 0 & 0 & 0\\
0 & 0 & 0 & 0
\end{array}\right),\,\,
\Lambda_{02}=
\left(\begin{array}{cccc}
0 & 0 & -1 & 0 \\
0 & 0 & 0 & 0 \\
1 & 0 & 0 & 0\\
0 & 0 & 0 & 0
\end{array}\right), \,\,
\Lambda_{03}=
\left(\begin{array}{cccc}
0 & 0 & 0 & -1 \\
0 & 0 & 0 & 0 \\
0 & 0 & 0 & 0\\
1 & 0 & 0 & 0
\end{array}\right), \,\,\nonumber\\
&& \Lambda_{12}=
\left(\begin{array}{cccc}
0 & 0 & 0 & 0 \\
0 & 0 & -1 & 0 \\
0 & -1 & 0 & 0\\
0 & 0 & 0 & 0
\end{array}\right), \,\,
\Lambda_{13}=
\left(\begin{array}{cccc}
0 & 0 & 0 & 0 \\
0 & 0 & 0 & -1 \\
0 & 0 & 0 & 0\\
0 & -1 & 0 & 0
\end{array}\right),\,\,
\Lambda_{23}=
\left(\begin{array}{cccc}
0 & 0 & 0 & 0 \\
0 & 0 & 0 & 0 \\
0 & 0 & 0 & -1\\
0 & 0 & -1 & 0
\end{array}\right)\,.
\eeq
The remaining $\Lambda_{\mu\nu}$ are obtained from the above
expressions by noting: $ \Lambda_{\mu\nu} = 
\Lambda_{\nu\mu}$. Parenthetically, we note that the S-transformed
$\lambda_{00}$ equals $\Lambda_{00}$ and is nothing but the standard
spacetime metric (for flat spacetime).

It can also be seen that  $\xi_\zeta(\vec p\,)$, for $\zeta=1,2,3$,
coincide with the solutions of Proca framework (and are divergence-less); 
whereas    
$\xi_4(\vec p\,)$, that gives the St\"uckelberg contribution to the 
propagator, lies outside the Proca framework:\footnote{We define the 
dragged Pauli-Lubanski
pseudovector, ${\widetilde{\W}}$, in a manner parallel to the introduction
of the dragged second Casimir operator.}

\begin{center}
\begin{tabular}{|c|c|c|c|c|}\hline
$\zeta$          & 
     $p_\mu \A\p  $  & 
                   ${\widetilde{\W}}^{(1/2,1/2)}\A\p $  & 
$\lambda_c$  & Remarks
 \\ \hline\hline
$1,2,3$ & $= 0$ & $\ne0$  & $ 2 $ & Proca Sector\\ 
\hline
$4$ & $\ne 0$ & $= 0$ & $ 0 $ & St\"uckelberg Sector\\ \hline		
\end{tabular}
\end{center}
In the above table we have introduced the $\lambda_c$ via the
equation:
\beq
\widetilde{C}_2^{(1/2,1/2)}{\mathcal A}\p = - \,\lambda_c\, m^2\,
{\mathcal A}\p\,.
\eeq

\section{Construction of spinor vector: The
massive gravitino}

We now wish to present the basis vectors for the representation
space defined by Eq. (\ref{rs}) in a language
which is widely used \cite{Moroi}. This would allow the present analysis
to be more readily available, and also bring out the
relevant similarities and differences with the 
framework of Rarita and Schwinger \cite{RS}.

In writing down the basis spinor-vectors, 
we  will use the fact that in  the $(1/2,1/2)$ 
representation space the charge conjugation is implemented by  
\beq
{\mathcal C}^{(1/2,1/2)}:\quad{\mathcal A}\p \rightarrow
\left[{\mathcal A}\p\right]^\ast\,. 
\eeq
In  the $(1/2,0)\oplus(0,1/2)$
representation space the charge conjugation operator is
$
{\mathcal C}^{(1/2, 0)\oplus(0,1/2)}:\quad i  \gamma^2 K\,,
$
where $K$ complex conjugates the spinor to its right. Then, with the uniform
choice, 
\beq
\vartheta_1= -\vartheta_2, \label{vartheta}
\eeq 
we obtain:
\beq
{\mathcal C}^{(1/2, 0)\oplus(0,1/2)}:{\Bigg\{}\begin{array}{l}
  u_{+1/2}\p \rightarrow - v_{-1/2}\p\,,
 u_{-1/2}\p \rightarrow   v_{+1/2}\p\,,\\
  v_{+1/2}\p \rightarrow   u_{-1/2}\p \,,
 v_{-1/2}\p \rightarrow - u_{+1/2}\p\,.
\end{array}
\eeq

In the spirit outlined,  the massive gravitino lives in a 
space spanned by sixteen spinor-vectors defined in items
{\bf A}, {\bf B}, {\bf C} below:

\begin{enumerate}
\item[{\bf A.}]
Of these,
eight spinor-vectors have $\widetilde{C}_2$  -- but not $\vec J^{\,2}$ -- 
eigenvalues, $-\, \frac{15}{4}\,m^2 $. These can be 
further subdivided
into particle,
\beq
\psi_a^\mu\p:
\cases{
& $\psi^\mu_1\p = u_{+1/2} \p \otimes \A_1\p$\,,\cr
&$\psi^\mu_2\p = \sqrt{\frac{2}{3}}\,  u_{+1/2} \p \otimes \A_2\p
                 + \sqrt{\frac{1}{3}}\,  u_{-1/2} \p \otimes \A_1\p$\,
,\cr
& $\psi^\mu_3\p = \sqrt{\frac{1}{3}}\,  u_{+1/2} \p \otimes \A_3\p
                 + \sqrt{\frac{2}{3}}\,  u_{-1/2} \p \otimes \A_2\p$\,
,\cr
& $\psi^\mu_4\p = u_{-1/2} \p \otimes \A_3\p$\,,\cr }\nonumber
\eeq
and antiparticle sectors:
\beq
[\psi_a^\mu\p]^{\mathcal C}:
\cases{
& $\psi^\mu_5\p = -v_{-1/2} \p \otimes \left[\A_1\p\right]^\ast$\,,\cr
& $\psi^\mu_6\p = - \sqrt{\frac{2}{3}}\,  v_{-1/2} \p \otimes \left[\A_2\p
\right]^\ast
                 + \sqrt{\frac{1}{3}}\,  v_{+1/2} \p \otimes \left[\A_1\p
\right]^\ast$
\,,\cr
& $\psi^\mu_7\p = - \sqrt{\frac{1}{3}}\,  v_{-1/2} \p \otimes
 \left[\A_3\p\right]^\ast
                 + \sqrt{\frac{2}{3}}\,  v_{+1/2} \p \otimes 
\left[\A_2\p\right]^\ast$
\,,\cr
& $\psi^\mu_8\p = v_{+1/2} \p \otimes \left[\A_3\p\right]^\ast$\,.\cr}
\nonumber
\eeq 
Here,
$[\psi_\tau^\mu\p]^{\mathcal C} = {\mathcal  C}^{(1/2,0)\oplus(0,1/2)}
\otimes{\mathcal  C}^{(1/2,1/2)} \,\psi_\tau^\mu\p$, 
$\tau=a,b,c$.

\item[{\bf B.}]
Four spinor-vectors have $\widetilde{C}_2$ -- but not $\vec J^{\,2}$ --  
eigenvalues, $-\, \frac{3}{4}\,m^2 $: 
\beq
\psi_b^\mu\p:
\cases{
& $\psi^\mu_9\p = \sqrt{\frac{2}{3}}\,  u_{-1/2} \p \otimes \A_1\p
                 - \sqrt{\frac{1}{3}}\,  u_{+1/2} \p \otimes \A_2\p$\,, 
\cr
& $\psi^\mu_{10}\p = \sqrt{\frac{1}{3}}\,  u_{-1/2} \p \otimes \A_2\p
                 - \sqrt{\frac{2}{3}}\,  u_{+1/2} \p \otimes \A_3\p$\,,
\cr}\nonumber
\eeq
\beq
[\psi_b^\mu\p]^{\mathcal C}:
\cases{
& $\psi^\mu_{11}\p = \sqrt{\frac{2}{3}}\,  v_{+1/2} \p \otimes 
\left[\A_1\p\right]^\ast
                 + \sqrt{\frac{1}{3}}\,  v_{-1/2} \p \otimes 
\left[\A_2\p\right]^\ast$\,,
\cr
& $\psi^\mu_{12}\p = \sqrt{\frac{1}{3}}\,  v_{+1/2} \p 
\otimes \left[\A_2\p\right]^\ast
                 + \sqrt{\frac{2}{3}}\,  v_{-1/2} \p \otimes 
\left[\A_3\p\right]^\ast$\,.\cr}\nonumber
\eeq

\item[{\bf C.}]
Another set of four spinor-vectors with $\widetilde{C}_2$  -- 
but not $\vec J^{\,2}$ --  
eigenvalues,  $-\, \frac{3}{4}\,m^2 $:  
\beq
\psi_c^\mu\p:
 \cases{ &
  $\psi^\mu_{13}\p = u_{+1/2} \p \otimes \A_4\p$\,,\cr
& $\psi^\mu_{14}\p = u_{-1/2} \p \otimes \A_4\p$\,,
\cr} \nonumber
\eeq

\beq
[\psi_c^\mu\p]^{\mathcal C}:
\cases{
& $\psi^\mu_{15}\p = - v_{-1/2} \p 
\otimes \left[\A_4\p\right]^\ast$\,,\cr
& $\psi^\mu_{16}\p = v_{+1/2} \p \otimes \left[\A_4\p\right]^\ast$
\,.\cr}
\nonumber
\eeq
\end{enumerate}

We have evaluated $\gamma_\mu\psi^\mu\p$,  $p_\mu\psi^\mu\p$, and 
$\widetilde{\mathcal W}^{(1/2,1/2)}_\mu \psi^\mu\p$, 
for all of the above sixteen 
spinor vectors. The $p_\mu\psi^\mu\p$, when transformed to the configuration
space, tests the divergence of $\psi^\mu(x)$.

For $\eta=1,4,5,8$, 
 $\gamma_\mu\psi^\mu\p$ identically vanishes. Requiring it
to vanish for $\eta=2,3,6,7$ results in: 
\begin{enumerate}
\item
$
\vartheta_1=\vartheta_2.
$
When combined with Eq. (\ref{vartheta}) this implies,
$
\vartheta_1= 0 = \vartheta_2
$.
As a consequence, the global phase factors $\vartheta_1$ and 
$\vartheta_2$ that appear in Eqs. (\ref{h}), (\ref{uv}), and 
(\ref{xi}) are not entirely free. 

\item
$E^2=\vert \vec p\,\vert^2 + m^2$
\end{enumerate}  

The $\tau=b,c$ sectors, if (wrongly) imposed with the 
vanishing of, $\gamma_\mu\psi^\mu\p$
and $p_\mu\psi^\mu\p$,  results in kinematically acausal 
dispersion relation (i.e., in $E^2\ne\vert \vec p\,\vert^2 + m^2$).
This  could be the source of the well-known 
problems of the Rarita-Schwinger framework 
as noted in works 
of Johnson and Sudarshan \cite{js}, and those of Velo and Zwanziger 
\cite{vz}. In this context one may wish to recall
that interactions can induce transitions between
different $\tau$ sectors.

The analysis for all the $\tau$ sectors of the
$\psi^\mu\p$ can be summarized in the following
table:

\begin{center}
\begin{tabular}{|c|c|c|c|c|c|}\hline
$\tau$          & 
     $p_\mu \psi^\mu\p  $  & 
                   $\gamma_\mu\psi^\mu\p $  & 
$\widetilde{{\mathcal W}}^{(1/2,1/2)}_\mu \psi^\mu\p$ & $ \lambda_c $ & Remarks
 \\ \hline\hline
$a$ & $= 0$ & $=0$  & $\ne 0$ & $\frac{15}{4}$ & Rarita-Schwinger Sector\\ 
\hline
$b$ & $= 0$ & $\ne 0$ & $\ne 0$ & $\frac{3}{4}$& {}\\ \hline	
$c$ & $\ne 0$ &
 $\ne 0$ & $=0$& $\frac{3}{4}$&{} \\ \hline	
\end{tabular}
\end{center}

The table clearly illustrates that there is no particular reason --
except  (the unjustified) 
insistence that each particle of nature be associated with a 
definite spin  -- 
to favor one $\tau$ sector over the other.
Each of the $\tau$ sectors
is endowed with specific properties. 
The Rarita-Schwinger sector 
has no more, or no less, physical significance
than the other two sectors. While, for instance, the 
Rarita-Schwinger sector can be characterized by vanishing of the 
   $p_\mu \psi^\mu\p  $  and 
                   $\gamma_\mu\psi^\mu\p $; the $\tau=c$ sector
is uniquely characterized by vanishing of 
$\widetilde{{\mathcal W}}^{(1/2,1/2)}_\mu \psi^\mu\p$. The  $\tau=b$ sector
allows for  vanishing of $p_\mu \psi^\mu\p  $ only. 

Except for the rest frame, the $\psi^\mu_\tau\p$, in general, 
are not eigenstates
of the $\vec J^{\,2}$ for representation space  (\ref{rs}).  
Instead, the three $\tau$ sectors of the representation space
under consideration 
correspond to the following inertial-frame independent values of the 
associated dragged second  Casimir invariant:
\beq
\widetilde{C}_2^{[(1/2,0)\oplus(0,1/2)]\otimes(1/2,1/2)} \psi_\tau^\mu \p=
-\, m^2\, \lambda_c \,\psi_\tau^\mu \p\,.
\eeq
For each of the $\tau$ sectors, 
the $\lambda_c$ are given in the table above.
Stated differently,
the $\tau=b,c$ sectors do not carry spin one half. Similarly, 
the, $\tau=a$, sector is  not a spin three half sector. The
consequence is that the  
$\tau=b,c$ sector, in particular, should not be treated as
a Dirac representation space. 
The correct wave equation for $\psi^\mu\p$ is:
\beq
\left[\left(\Lambda_{\mu\nu} 
 p^\mu p ^\nu \pm m^2 I_4\right)\otimes
\left(\gamma_\mu p^\mu \pm m I_4\right)\right]\psi^\epsilon\p =0\,.
\eeq

In the standard Rarita-Schwinger
framework $\partial_\mu\psi^\mu(x)$ and
$\gamma_5 \gamma_\mu\psi^\mu(x)$ do indeed behave as Dirac spinors,
and do indeed satisfy the Dirac equation. However, they
are not identical to the $\tau=b,c$ sectors (which do not
carry a characterization in terms of spin one half). 
If one (mistakenly) makes this identification, and sets 
$\partial_\mu\psi^\mu(x)$ and
$\gamma_5 \gamma_\mu\psi^\mu(x)$ to zero, one introduces an 
element of kinematic acausality.

\section{Interpretation}

If one is to respect the mathematical completeness of the 
spinor-vector representation space associated with $\psi^\mu(x)$, the
Rarita-Schwinger framework cannot be considered to describe
the full physical content of the representation space associated with
a massive gravitino. 
This circumstance is akin to  Dirac's
observation that a part of a representation space [which would have
violated the completeness of the $(1/2,0)\oplus (0,1/2)$] 
cannot be ``projected out'' without introducing certain
mathematical inconsistencies, and loosing its physical content (i.e. 
antiparticle, or particles).\footnote{Similar remarks apply to
Majorana construct in the $(1/2,0)\oplus (0,1/2)$ representation 
\cite{dva_maj}.}
Further, the same qualitative remarks
apply to the $(1/2,1/2)$ representation space when in the
Proca framework one only confines to the divergence-less vectors.
The ``projecting out'' of the divergence-full vector, throws away 
the St\"uckelberg contribution to the 
propagator, and in addition leaves the $(1/2,1/2)$
representation space
 mathematically incomplete.
Now, we suggest that for the representation space defined by
Eq. (\ref{rs}), one needs to consider all three $\tau$ sectors 
of $\psi^\mu(x)$ 
as physical, and necessary for its mathematical consistency. 
The suggested framework already carries consistency
with the known data on the $N$ and $\Delta$ resonances, and asks
that massive gravitino be considered  as an object that is better described by 
the eigenvalues of the dragged 
second Casimir operator. In its rest frame it 
is endowed with a spin three half, 
and two spin half, components.  
A spin measurement for unpolarized ensemble of  massive gravitinos
at rest would yield the results $3/2$ with probability one half, and
$1/2$ with probability one half. The latter probability
is distributed uniformly, i.e. as
one quarter,  over each of  the, $\tau=b$, and, $\tau=c$, sectors.

\newpage
\begin{figure}
\begin{center}
\vskip -3 cm
\includegraphics*[width=11cm]{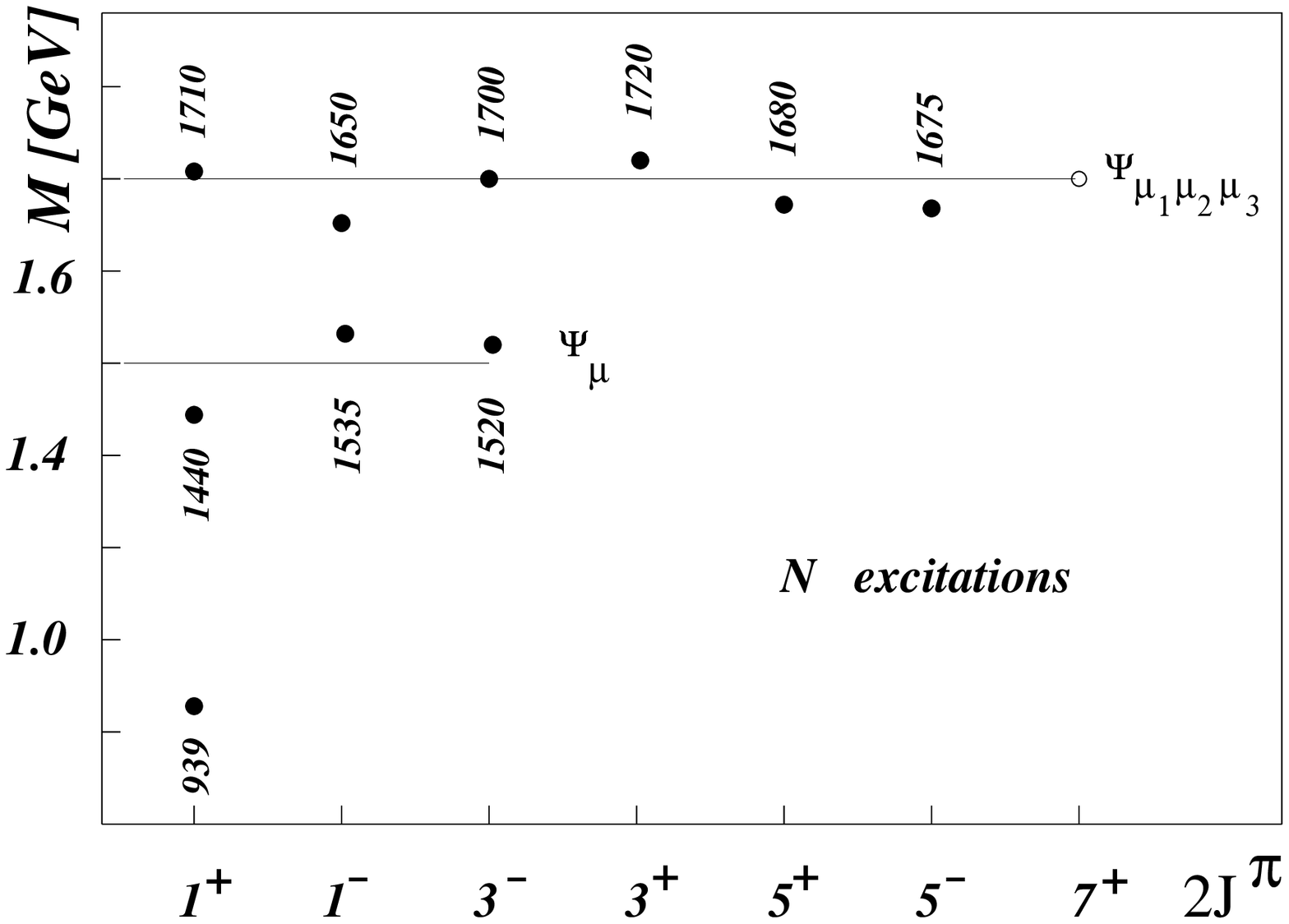}
\vskip -8cm
\includegraphics*[width=11cm]{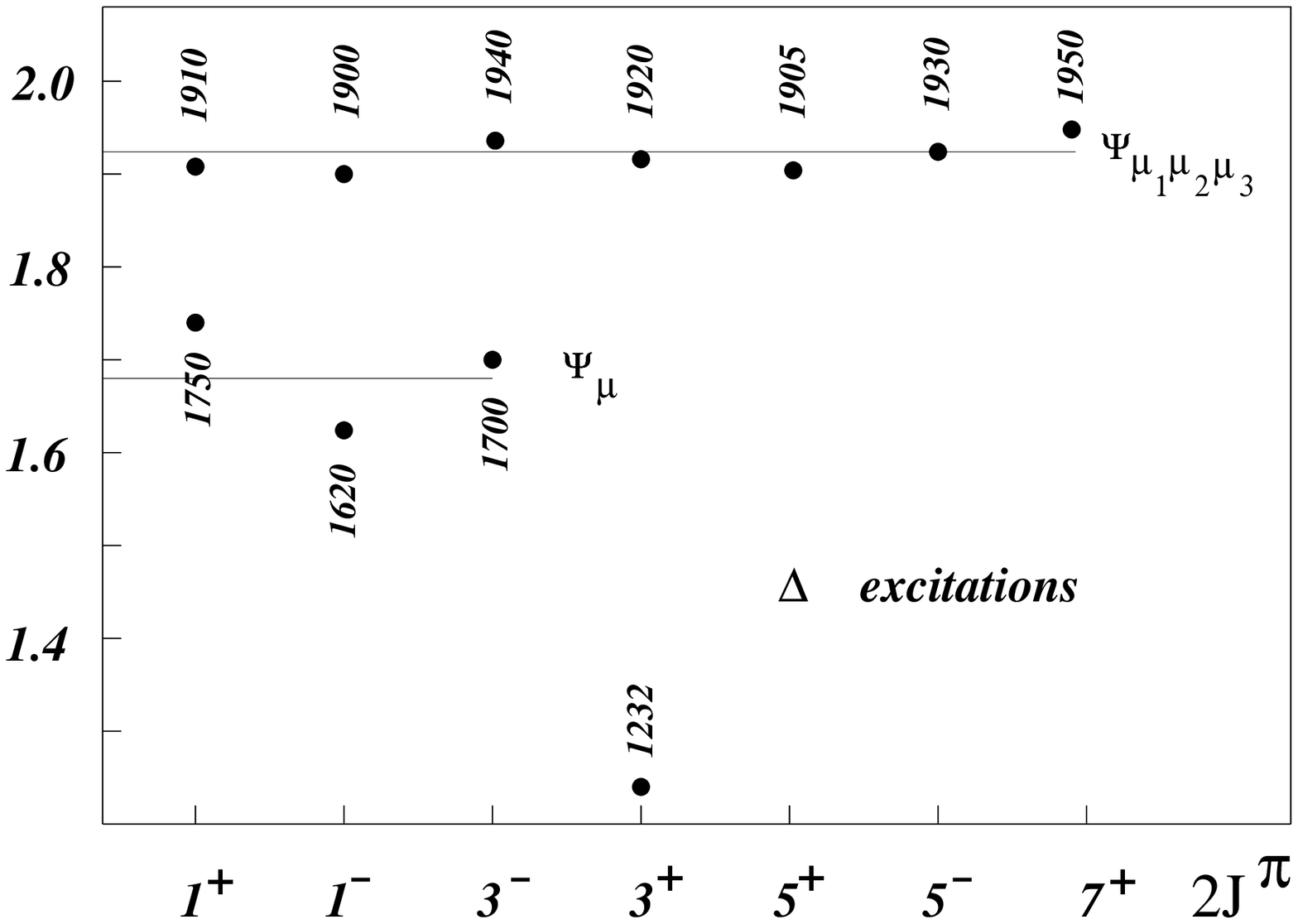}
\end{center}
\caption{Summary of the data on $N$ and $\Delta$ resonances.
The breaking of the mass degeneracy for each of the clusters
at about $5\%$ may in fact be an artifact of the data analysis,
as has been suggested by H\"ohler \cite{gh}.
The filled circles represent known resonances, while 
the sole empty circle corresponds to a prediction. 
}
\end{figure}


\begin{thebibliography}{000}


\bibitem{sw}
S. Weinberg, {\em The quantum
theory of fields, Vol. I and III\/} (Cambridge University Press,
Cambridge, 1995 and 2000).


\bibitem{ak}
D. V. Ahluwalia, M. Kirchbach, Mod. Phys. Lett. 
{\bf A16} (2001) 1377.

\bibitem{v}
M. Veltman, Int. J. Mod. Phys. {\bf A15} (2000) 4557.

\bibitem{mk}
M. Kirchbach, Mod. Phys. Lett. {\bf A12} (1997) 2373;\\ M. Kirchbach, 
Nucl. Phys.
{\bf A689} (2001) 157c.

\bibitem{gh}
G. H\"ohler, in {\em Pion-Nucleon Scattering\/} 
(Springer Publishers, Heidelberg, 1983), Landolt-B\"ornstein
Vol. I/9b2, Ed. H. Schopper.

\bibitem{pdg}
Particle Data Group, Eur. Phys. J,  {\bf C15} (2000) 1.

\bibitem{Moroi}
T. Moroi, Ph. D. thesis (Tohoku University, Japan, 1995), hep-ph/9503210.

\bibitem{RS}
W. Rarita, J. Schwinger, Phys. Rev. {\bf 60} (1941) 61.

\bibitem{js}
K. Johnson, E. C. G. Sudarshan, Ann. Phys. (N. Y.), {\bf 13}
(1961) 126.

\bibitem{vz}
G. Velo, D. Zwanziger,
Phys. Rev. {\bf 186} (1969) 1337.

\bibitem{dva_maj}
D. V. Ahluwalia,
Int. J. Mod. Phys. {\bf A11} (1996) 1855.

\end{thebibliography}
\end{document}